# Cause Identification of Electromagnetic Transient Events using Spatiotemporal Feature Learning

Iman Niazazari, Reza Jalilzadeh Hamidi, Hanif Livani, and Reza Arghandeh

*Abstract*—This paper presents a spatiotemporal unsupervised feature learning method for cause identification of electromagnetic transient events (EMTE) in power grids. The proposed method is formulated based on the availability of time-synchronized high-frequency measurement, and using the convolutional neural network (CNN) as the spatiotemporal feature representation along with softmax function. Despite the existing threshold-based, or energy-based events analysis methods, such as support vector machine (SVM), autoencoder, and tapered multi-layer perception (t-MLP) neural network, the proposed feature learning is carried out with respect to both time and space. The effectiveness of the proposed feature learning and the subsequent cause identification is validated through the EMTP simulation of different events such as line energization, capacitor bank energization, lightning, fault, and high-impedance fault in the IEEE 30-bus, and the real-time digital simulation (RTDS) of the WSCC 9-bus system.

## I. INTRODUCTION

### A. Problem Statement and Literature Review

Electromagnetic transient events (EMTE), such as faults, line energization, capacitor bank energization, and lightning occur intermittently in power systems and can result in momentary or permanent disturbances. These events can cause immediate or long-term damage to electrical equipment, such as transformers, circuit breakers, and capacitor banks. According to [1][1], 27% of the momentary events between years 2003 and 2008 have been due to lightning. Utilities without effective EMTE detection and classification tools may eventually face a permanent failure with significant damage to their electrical equipment.

EMTE analysis is widely studied in the literature. In [2], the wavelet transform is introduced for characterizing the electromagnetic transient behavior of switching events and faults in power grids. In [3], the authors propose a wavelet-based method for high impedance faults detection in transmission systems. The method uses a supervised feature learning method and only captures the temporal features of the measurements. A new wavelet entropy and the neural network-based technique for EMTE classification is proposed in [4]. In this paper, the wavelet energy entropy (WEE) and wavelet entropy weight (WEW) are used as the supervised features to a back-propagation neural network classifier. However, the classifications framework only deals with measurement signals transmitted from only one location and multiple streaming measurements are not considered. In [5], a transient overvoltage localization method is proposed for two types of events, capacitor bank energization, and ground faults, using the wavelet packet decomposition as the feature learning method, and a neural network as the event classifier. In [6], a method based on the wavelet transform and hybrid principal component analysis is proposed to classify and locate the switched capacitor bank events. A new two-steps technique for switched capacitor banks classification is presented in [7]. In the first step, the exact instant of switching is detected. In the second step, the distance between the switched capacitor and the point of monitoring is then calculated using the linear circuit theory. The disadvantage of the proposed method is that the accuracy of the estimated distance relies on the accuracy of the measured short circuit current. In [8], a time-frequency analysis of voltage measurements is proposed for lightning detection in cables. The proposed method only relies on temporal features from only one location. In [9], EMTE classification is performed using autoencoder as an unsupervised feature learning algorithm, and the softmax as the event classifier. However, it considers data stream from only one single measurement.

The use of the convolutional neural network (CNN) for cause identification of events in power systems is presented in several works [10]-[13]. However, the proposed methods use only time-synchronized measurement from one location or several unsynchronized measurements. In [14], CNN is utilized for event classifications based on the phasor measurement units (PMUs) data. The events include generator tripping, line tripping, load disconnection, and inter-area oscillation. The measurements from one substation are transformed into 2D inputs that results in temporal learning. Reference [15] proposes a fault detection and classification method using sparse autoencoder and CNN. The time-synchronized voltage and current data are obtained from only one measurement device for temporal feature extraction and event classification.

As briefly discussed above, most of EMTE analysis methods and other power quality classification methods, such as the ones in [16]-[21], do not consider the spatiotemporal features of measurements data or their time-frequency components, such as wavelet coefficients. The aforementioned methods are mostly designed based on either spatial or temporal features of measurements. In such methods, the

I. Niazazari and H. Livani are with the Department of Electrical and Biomedical Engineering, University of Nevada, Reno, NV, USA 89557 (e-mail: niazazari@nevada.unr.edu and hlivani@ieee.org).
R.J. Hamidi is with the Department of Electrical Engineering, Arkansas Tech University, Russellville, AR, USA 72801 (email: rhamidi@atu.edu)
R. Arghandeh is with Department of Computing, Mathematics and Physics, Western Norway University of Applied Sciences, Bergen, Norway (email: arghandehr@gmail.com)

features are obtained for a number of fixed intervals in time or space using mathematical operations, such as mean, summation, Euclidian norm, infinite norm, and entropy, on the short-time Fourier transform (STFT) or the discrete wavelet transform (DWT). However, events in power systems can have very similar spatial or temporal features, such as the total wavelet energy as a spatial feature or total duration of an event as the temporal feature. Therefore, EMTE cause identification may not be easily carried out by only monitoring the relay voltage and current outputs or the peak values or duration of the time-frequency results. However, different events create unique spatiotemporal patterns on the measurements or on the extracted time-frequency components that can be exploited for effective event analysis. In spatial or temporal feature learning methods, the spatiotemporal correlation of the measurements cannot be extracted simultaneously, and we need to mix the features in time and space separately by stacking the features from different locations in early fusion or stacking the features in late fusion [22].

*B. Contribution*

To the best of the authors' knowledge, the use of spatiotemporal unsupervised feature learning of wide-area high frequency (HF) measurements for cause analysis of power system events has not been adequately addressed in the literature. This paper presents a data-driven cause identification of very-fast transient events in power systems based on CNN[23]. The results of this work can benefit the utilities to increase the situational awareness of very fast EMTE such as faults and high-impedance faults, and enhance end-of-the-year power quality assessment and cause analysis reports. The contributions of the paper are summarized as follows:

- From the methodology point of view, since the measurements form network have a very spatiotemporal dependency on each other, deploying a spatiotemporal feature learning is suggested. We propose an unsupervised feature learning method based on CNN to capture the spatiotemporal correlation of the wide-area HF measurements by convolving several filters through the stream of measurements. CNN benefits from the strong dependency existing among neighboring pixels in the input data. The proposed unsupervised spatiotemporal feature learning method will result in better cause identification of events compared to existing spatial or temporal feature learning methods. The methods mentioned in other papers either did not take advantage of the spatiotemporal of measurements or they did not consider the possibility of multiple measurements at the same time.
- From the application point of view, important events from operators' perspective, such as fault, high-impedance fault, lightning, line energization, and capacitor bank energization, with very similar spatial or temporal features are considered. These events are very fast with a very small life span and complex behavior compared to the majority of other events such as generating units disturbances. Consequently, cause identification of such fast events is a very challenging problem which is solved using the proposed CNN-based method. Moreover, new use cases are developed for advanced wide-area HF time-synchronized measurements that stream data from multiple locations.

The proposed method is validated using realistic EMTE data produced by a real-time digital simulator (RTDS) for three events including fault, line energization, and capacitor bank energization. The rest of this paper is organized as follows: Section II introduces the CNN-based spatiotemporal feature learning and the cause identification framework. Events descriptions are presented in Section III. The RTDS and EMTP simulation test cases, results, and discussions are presented in Section IV followed by the conclusion in Section V.

## II. SPATIOTEMPORAL UNSUPERVISED FEATURE LEARNING FOR CAUSE IDENTIFICATION

This paper proposes the use of the convolutional neural network (CNN) to learn the spatiotemporal features of the wavelet transform coefficients (WTCs) of HF voltage measurements in an unsupervised manner and fed the features to a supervised classifier for cause identification. CNN is a powerful tool that can be used for spatiotemporal unsupervised feature learning of data stream and the subsequent supervised cause identification of the underlying event. It has been applied to numerous applications, such as image classification, video classification, and sentence classification [24]-[27]. The architecture of the proposed EMTE cause identification method is illustrated in Fig. 1. In the following subsections, different blocks of the overall algorithm are presented.

*A. Spatiotemporal Data Input Block*

The HF voltage data are captured by time-synchronized transient recorders in different locations. The modal transformation is then applied to obtain the modal voltages, i.e., mode 1, mode 2, and mode 0. To consider different possibilities of input data to the spatiotemporal feature learning block, four different scenarios are considered.

*Case 1)* the original mode 1 voltages from multiple locations are transformed into a grayscale (2D) image.

*Case 2)* DWT is applied to the mode 1 voltage stream from multiple locations and the absolute value of the wavelet transform coefficients (|WTC|s) are converted into a grayscale (2D) image.

*Case 3)* the original mode 1, 2, and 0 are transformed into an RGB (3D) image.

*Case 4)* DWT is applied to the original mode 1, 2, and 0 voltages and the |WTC|s are converted into an RGB (3D) image.

DWT is a powerful tool for capturing time-frequency characteristics of measurements and has been extensively applied to power system application, including electromagnetic transient events and fault analysis. In this paper, DWT is applied to obtain the high and low-frequency components. For short and fast events, such as EMTE, Daubechies-4 (db4) is used to accurately extract the time-frequency information [28][30]. In this paper, we have also verified the superior performance of db4 in scale-1 to accurately capture the transient information and subsequently carry out the cause identification of events, compared to other well-known mother wavelets in different scales, such as db2, db8. The absolute values of the HF components (the detail of the DWT in scale 1) are calculated (|WTC|s) and then normalized with respect to their peaks, and

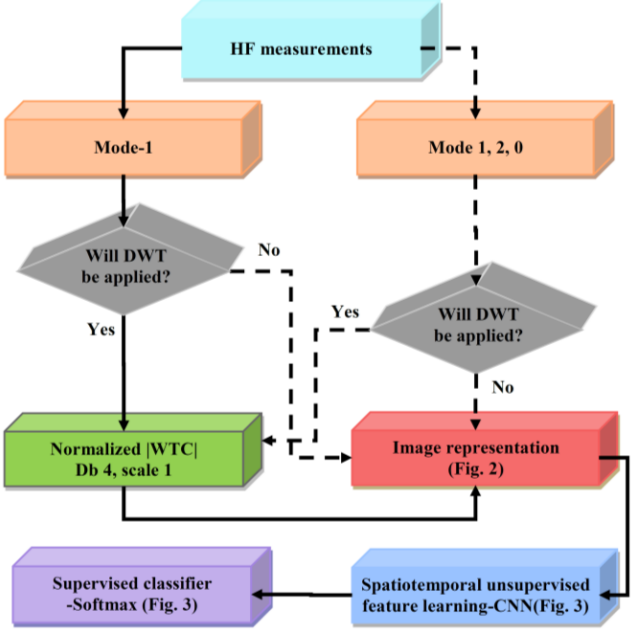

Fig. 1. EMTE root cause analysis flowchart

then used as the input to the proposed CNN-based framework.

The input to the image representation block is converted to the image where each column corresponds to the time, and each row corresponds to the measurement location. Fig. 2 shows the image conversion for case 2 where the |WTC|s are converted to a greyscale image, where the absolute value of the |WTC|s in time and location are represented as a matrix using the following conversion

$$Grayscale = \frac{|WTC_{L,S}| - |WTC_{L,min}|}{|WTC_{L,max}| - |WTC_{L,min}|} \times 255 \quad (1)$$

where $|WTC_{L,S}|$ is the absolute value of the wavelet transform coefficient at sample $S$ and location $L$, $|WTC_{L,min}|$ is minimum absolute value of wavelet transform coefficient at location $L$, and $|WTC_{L,max}|$ is maximum absolute value of wavelet transform coefficient at location $L$.

### B. Event Cause Classification Block

CNN is a specific class of deep feed-forward artificial neural networks that has been utilized in many scientific fields such as image processing. The use of CNN for cause identification of events is beneficial as it is a spatiotemporal unsupervised feature learning approach that is proven to outperform the supervised feature learning methods. A CNN is composed of an input, an output layer, and several middles layers known as hidden layers.

The hidden layers of a CNN are composed of convolutional layers, pooling layers, and fully connected layers which are explained in the following sections. The structure of CNN is shown in Fig. 3.

*1) Convolutional layer*

A convolutional layer uses a convolution operation to pass the input to the next layer. A convolutional layer comprises neurons that are connected to sub-regions of the input data. The features of the sub-regions inside the measurement matrix are learned through the convolution layer. The size and number of these regions can be specified during the convolutional layers'

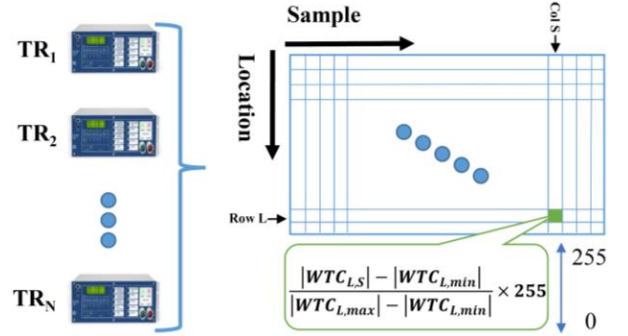

Fig. 2. Spatiotemporal representation of the |WTC| using grayscale selection.

Suppose layer $l$ is a convolutional layer. The output of the layer $l$ consists of $N_1^{(l-1)}$ feature maps from the previous layer, each with size $N_2^{(l-1)} \times N_3^{(l-1)}$. The output of layer $l$ is composed of $N_1^{(l)}$ feature maps with size of $N_2^{(l)} \times N_3^{(l)}$. The $i^{th}$ feature map in layer $l$, denoted as $Y_i^{(l)}$, is computed as

$$Y_i^{(l)} = B_i^{(l)} + \sum_{j=1}^{N_1^{(l-1)}} K_{i,j}^{(l)} \times Y_j^{(l-1)} \quad (2)$$

where $B_i^{(l)}$ is a bias matrix and $K_{i,j}^{(l)}$ is the filter connecting the $j^{th}$ feature map in layer $(l-1)$ with the $i^{th}$ feature map in layer $l$ [31].

*2) Pooling layer*

The output neurons of the convolutional layer are clustered together and combined into one neuron to speed up the feature learning process. This process occurs in the pooling layer. There are usually two types of pooling; max pooling and average pooling. In the max pooling, the pooling value is the maximum value from each of a cluster of neurons at the previous layer. Similarly, in the average pooling, the pooling value is the average value from each of a cluster of neurons at the preceding layer.

*3) Fully-connected layer*

Neurons in a fully-connected layer have full connections to all neurons at the previous layer, as in traditional neural network-based methods, such as autoencoders. In the CNN method, the output of fully-connected layer is fed to a classifier, such as a *softmax* function. The fully connected layer represents the learned spatiotemporal features as a vector. Further technical details can be found in [32].

*4) Softmax Classifier*

The softmax classifier is used as a supervised classifier to assign the learned features from the fully connected layer to their corresponding classes. Softmax function is a generalization of the logistic function with a multiclass probability distribution as opposed to a binary probability distribution. The softmax classifier expresses the probabilistic representation of each class as shown in (3).

$$P(y = i|z) = \frac{e^{(z^T w_{si} + b_{si})}}{\sum_{j=1}^{N} e^{(z^T w_{sj} + b_{sj})}} \quad i = 1, \dots, N \quad (3)$$

where z is the output vector of the fully connected layer in CNN, $y$ is the class label, $W_s \in R^{m \times c}$ is the weight vector, $b_s \in R^c$ is the bias vector, and $N$ is the total number of classes [33].

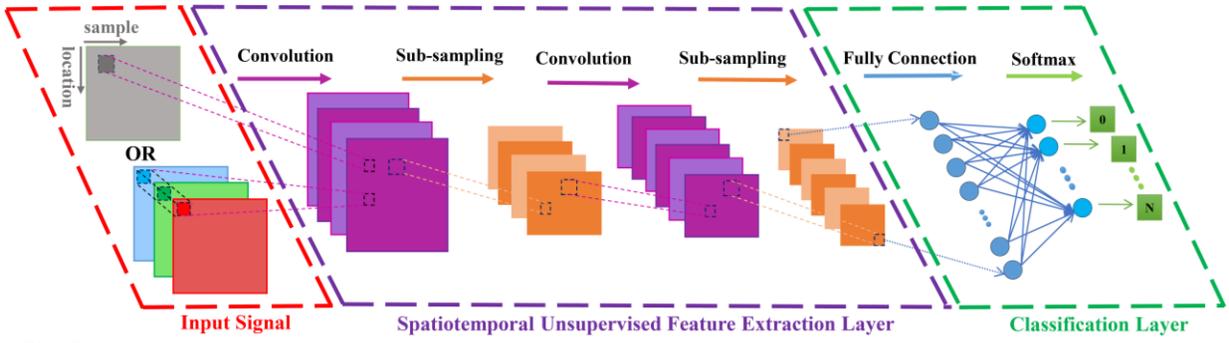
Fig.3 . CNN Structure

## III. Events Description and Use Cases

The data set is first created using the real-time digital simulator (RTDS) to resemble realistic EMTE. In addition, for an extensive study, a more comprehensive data set is created using the electromagnetic transients program (EMTP). In this paper, we have assumed that the events are detected using the robust algorithms proposed by the authors in [34],[35] .We will then use the measurements for the duration of two cycles after the occurrence of the events before any relay operates. This is determined heuristically based on extensive studies of different events and their impacts on the measurements. In this paper, five major EMTE classes are considered and described below. In this paper, we select these events to represent a range of fast disturbances with similar impacts on measurements, that frequently occur in power systems. However, there might be other types of events with similar patterns that we can consider them as additional classes, if the source of events are known to the system operators and the data sets can be labeled accordingly. In the case of unknown cause of events, we can combine them into one single class representing all the unknown events [36], and then utilize the proposed framework. The events cover a diverse set of conditions, such as type, location, inception angle, and fault resistance.

- *Class 1 (Line Energization):* Transmission line energization can be carried out using two strategies, unsynchronized operation of circuit breakers at both ends of a line, and synchronized operation of circuit breakers at both ends of a line which represents point-on-wave (PoW) switching strategy [37].
- *Class 2 (Capacitor Bank Energization):* Switching of capacitor banks create high magnitude and high-frequency transients due to the inrush current coming from the system sources [38].
- *Class 3 (Fault):* All types of single-phase, two-phase, and three phase faults with different fault resistance, fault inception angles, and locations are considered.
- *Class 4 (Lightning):* [only EMTP] This class is simulated based on the CIGRE standard lightning model that strikes different locations along each transmission line at every ten miles. To have a more comprehensive lightning event repository, two types of lightning strikes are simulated, namely, shielding failure representing the surging current of 30 kA and back flashover representing the currents of 100 kA [39].
- *Class 5 (High Impedance Fault (HIF)):* [only EMTP] HIFs usually occur because of unwanted electrical contact between a conductor and high impedance trees, or between a broken conductor and the ground [3].Since the fault current remains below the threshold of overcurrent relays, detecting this type of fault is a much more challenging task. The high impedance fault is simulated using the model given in [40].

The summary of events descriptions, a sample mode 1 voltage in EMTP, the corresponding |WTCs|, all the considered parameters, and the total number of simulated events are presented in Table I.

## IV. Results and Discussions

We first present the results of the WSCC 9-bus system using the RTDS, for three classes of line energization, capacitor bank energization, and fault. We validate the performance of the proposed method by comparing the results with support vector machine (SVM), autoencoder, and a non-spatiotemporal-based feedforward neural network classifiers. Finally, for a more extensive validation, all five events are simulated in the IEEE 30-bus system using the EMTP software. The validation of the proposed method is carried out with respect to different scenarios, such as number and location of measurements.

### A. RTDS Results

To validate the proposed methodology, the WSCC 9-bus system is simulated using the RTDS. Three software-in-the-loop TRs are considered at buses 4, 7, and 9. The simulation time step is 50 μs. Three events are simulated as Class 1 (transmission line energization, Class 2 (capacitor bank energization), and Class 3 (fault). Classes 4 and 5 are required to be simulated using the small time step blocks (1.4 $\mu s$ time step). Because of the limited processing capacity of the UNR's RTDS for small time step simulations, Classes 4 and 5 are not considered.

Fig. 4 shows the RTDS implementation of the WSCC 9-bus system. The proposed CNN-based method is then used for the cause identification of three EMTE. The initial weights for all the CNN layers are set with a normal distribution with zero mean and standard deviation of 0.01, and the initial value of biases are zero.

Numerous numbers of training and evaluation scenarios are carried out to find the best CNN parameters, i.e., the number of filters, convolutional layers, fully connected (FC) layers, and size of the filters. The optimal CNN structure is as follows: one convolutional layer with ten 2×20 filters, the stride of 1×1, maximum pooling of 1×2, and one fully connected layer. The classifier is trained on a single CPU using the stochastic gradient descent method with momentum with the number epoch of 40, mini-batch size of 128, the initial learning rate of 0.0001, and momentum of 0.9. Therefore, in all the following discussions, we use these selected CNN parameters. In addition, for the evaluation, 80% of the data set is used for training, and the rest is used for testing.

TABLE I
EVENTS DESCRIPTIONS SUMMARY

| Event | Description | Mode 1 Measurements (in EMTP) | \|WTC\|s (in EMTP) | EMTP Simulations | RTDS Simulations |
|---|---|---|---|---|---|
| **Transmission Line Energization** | Closing the transmission lines (synchronizes, unsynchronized) | 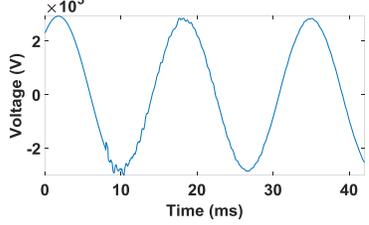 | 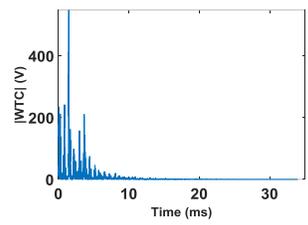 | Unsynchronize: 16 switching instances 82 switches / Synchronized: 16 switching instances 41 lines 4 delays / Total=3936 | Unsynchronize: 16 switching instances 12 switches / Synchronized: 16 switching instances 6 lines 4 delays / Total=576 |
| **Capacitor Bank Energization Inrush** | Switching the capacitor bank | 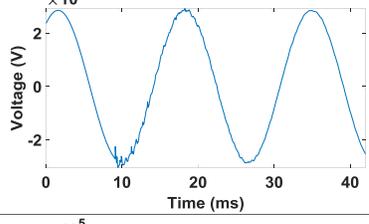 | 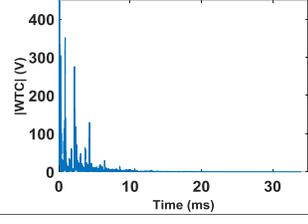 | 11 cap 8 values per cap 16 switching instances / Total=1408 | 3 cap 4 values per cap 16 switching instances / Total=192 |
| **Fault** | Different types of failures in the network | 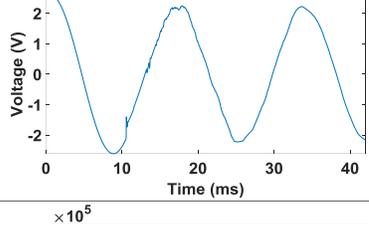 | 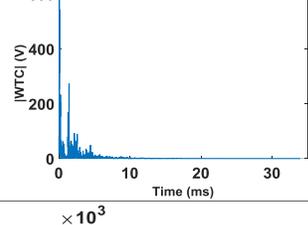 | 1G, 2G, 3G Faults per 10 mi $R_f$= 0.1, 5 ohm 3 incpetion angle / Total=11088 | 1G, 2G, 3G Faults per 20% $R_f$= 0.1, 5 ohm 4 incpetion angle / Total=576 |
| **Lightning** | Lightning strikes (shielding failure, back flashover) | 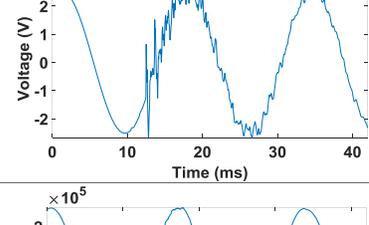 | 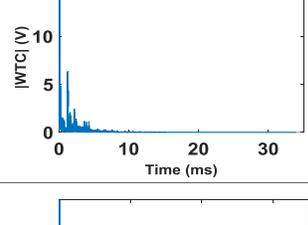 | Shielding failure: current 30 kA strike per 10 mi / Back flashover: current 100 kA strike per 10 mi / Total=1232 | N/A |
| **High Impedance Fault (HIF)** | Unwanted electrical contact between primary conductors and the vegetation | 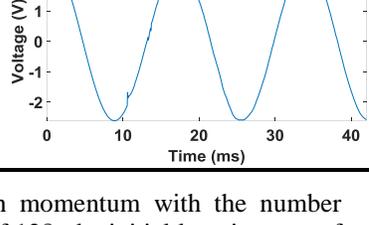 | 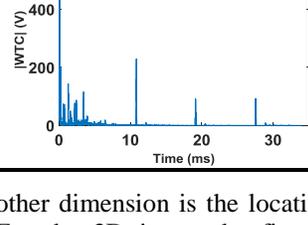 | Faults per 10 mi 6 $R_A$ and $R_B$ 3 inception angle / Total=11088 | N/A |

To show the superiority of the proposed spatiotemporal method, the cause identification accuracy is compared with three other state-of-the-art methods, which are PCA+SVM, autoencoder, and tapered multilayer perception neural network (t-MLP) [41]. The accuracy is the percentage of number of correctly predicted events over the total number of predictions. Moreover, rows in Table II show the accuracies for four different types of inputs (i.e., cases defined in Section II). In Table II, D stands for dimension and W stands for wavelet transform on modal voltage measurements at different locations. In the case of 2D input, one dimension is time and the other dimension is the location of the voltage measurements. For the 3D input, the first dimension is time, the second dimension is measurement location, and the third dimension is the mode of the voltage measurement (mode 0, 1 and 2). Among the four different variations of input in Table II, the 2D+W provides the highest accuracy. Among different methods, our proposed method significantly outperforms the others. Another interesting observation is that applying DWT increases the accuracy of the method. Moreover, the 2D+W is slightly better than the 3D+W which means adding more dimension does not necessarily lead to higher accuracy.

To further validate the performance of the proposed method, the accuracy across different iterations for four different input cases is shown in Fig. 5. As it can be seen in the figure, the proposed 2D+W case consistently outperforms the other type of inputs. Therefore, from this point forward, we stick to this case and use it for further validation.

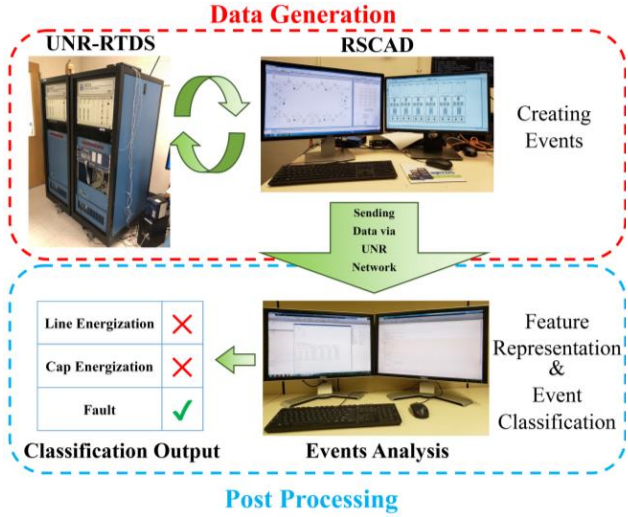

Fig. 4. RTDS implementation of the WSCC 9-bus system

TABLE II
CAUSE IDENTIFICATION ACCURACY COMPARISON BETWEEN DIFFERENT INPUTS AND DIFFERENT METHODS (D: DIMENSION, W: WAVELET)

| Input | PCA+SVM Accuracy (%) | Autoencoder Accuracy (%) | t-MLP Accuracy (%) | CNN Accuracy (%) |
|---|---|---|---|---|
| 2D | 78.35 | 62.78 | 62.69 | 94.40 |
| **2D+W** | **83.20** | **74.81** | **68.66** | **98.52** |
| 3D | 63.05 | 67.29 | 84.33 | 95.52 |
| 3D+W | 75 | 71.80 | 80.60 | 97.39 |

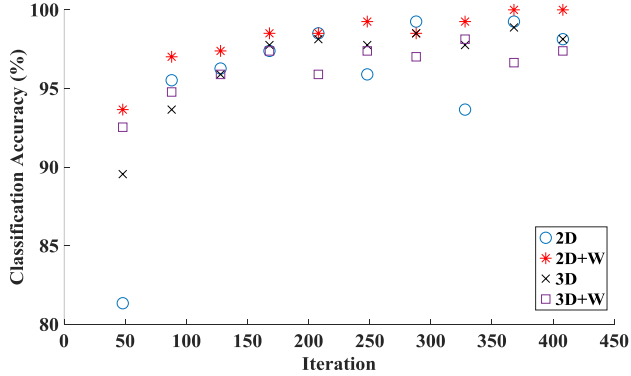

Fig. 5. Comparison between different inputs of data using CNN

## B. EMTP Results

The proposed methodology is further validated using the EMTP simulation results of a larger use case, the IEEE 30-bus system. All the five events in Section III are simulated based on the parameters mentioned in Table I. It is assumed that the synchronized TRs are installed on several buses. The frequency-dependent line models with the default EMTP parameters are used for all transmission lines. The CNN parameters are the same as the RTDS use case, except the optimum size of the convolution filters that is 5×100 which is obtained with extensive simulations. In addition, the training is carried out with number epoch of four, mini-batch size of 128, initial learning rate of 0.0001, and momentum of 0.9.

### 1) Impact of Number of Measurement Units

Synchronized TRs are among the most expensive advanced monitoring devices installed in power systems [42]-[44]. In this regard, the knowledge on the accuracy of the proposed method with a different number of TRs installed in the network will be of great interest. To carry out such sensitivity analysis, it is assumed that TRs stream time-synchronized voltage measurements from certain nodes. To assess how the number of TRs affect the overall accuracy, the process is carried out four times with two (at bus 6 and 10), five (at bus 6, 10, 4, 28, and 8), and ten (at bus 6, 10, 4, 28, 8, 22, 21, 9, 12, and 3) TRs installed in the network. As it is observed in Fig. 6, with the increase in the number of TRs, the overall accuracy increases. This is due to the fact that an increase in the number of installed TRs leads to transferring more HF data from different locations of the grid, which consequently enhances the feature learning and cause identification performance. However, with even two installed TRs, the accuracy is in the acceptable range.

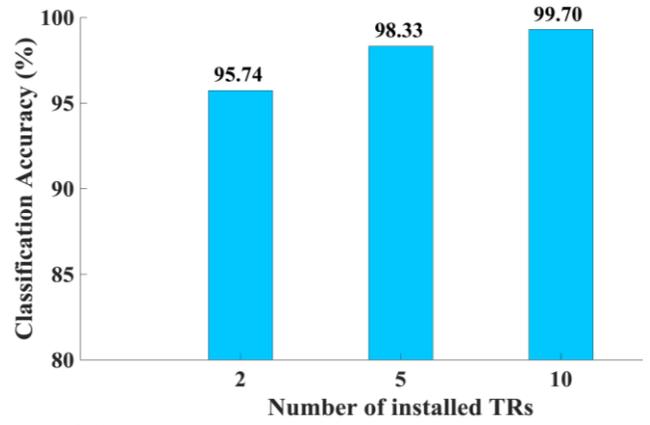

Fig. 6. Cause identification accuracies with respect to the number of TRs in the IEEE 30-bus system

### 2) Impact of Measurement Units Placement

The impact of TRs placement on the accuracy is discussed in this section. It is beneficial for utilities to install such expensive TRs at certain substations to enhance the cause identification accuracy. In [45], three different metrics, namely electrical connectivity, electrical centrality, and electrical node significance, are utilized to determine the buses with strong or weak electrical connections with the remaining of the grid. These buses have some impact on the spatiotemporal correlation of the nodal voltages.

In this paper, the electrical centrality index is deployed to find the dominant buses for installing synchronized TRs. Based on this index, buses with relatively small electrical centrality index (e.g., radial buses) are weak in terms of electrical connectivity and vice versa. The voltages at the buses with strong connectivity have a strong temporal correlation. After applying the electrical centrality index to the system, the ten dominant sequence of the buses is A={6, 10, 4, 28, 8, 22, 21, 9, 12, 3} as shown in Table III. This table compares the accuracies in five different cases. In the first scenario, TRs are installed at the obtained dominant buses (A). In the other four scenarios, TRs are randomly placed in the grid as shown in Table III. As it can be seen, the accuracy for the first case, where the electrical connectivity index is used for TR placements, is slightly enhanced compared to the other scenarios.

TABLE III
OVERALL CAUSE IDENTIFICATION ACCURACIES WITH DIFFERENT
LOCATION OF TRs IN THE IEEE 30-BUS SYSTEM

| TRs locations | Classification Accuracy (%) |
|---|---|
| **A={6, 10, 4, 28, 8, 22, 21, 9, 12, 3}** | **99.70** |
| B={19, 3, 10, 7, 30, 26, 5, 9, 14, 6} | 99.04 |
| C={8, 27, 9, 24, 17, 18, 4, 5, 28, 30} | 99.18 |
| D={2, 18, 30, 28, 14, 6, 7, 13, 19, 9} | 99.23 |
| E={13, 23, 28, 4, 30, 14, 19, 11, 27, 6} | 89.99 |

TABLE IV
CAUSE IDENTIFICATION ACCURACY FOR THE IEEE 30-BUS SYSTEM USING
ENERGY-BASED, NON-SPATIOTEMPORAL FEEDFORWARD NN AND CNN-BASED
SPATIOTEMPORAL METHOD

| Method | | ACC | PRE | REC | F1 | FPR |
|---|---|---|---|---|---|---|
| Energy-based | PCA+SVM | 74.53 | 69.95 | 74.68 | 72.24 | 6.62 |
| | Autoencoder | 81.32 | 66.75 | 69.93 | 68.30 | 5.44 |
| Non-Spatiotemporal | t-MLP | 92.40 | N/A | 74.45 | N/A | 2.37 |
| Spatiotemporal | CNN | 99.70 | 99.63 | 99.24 | 99.43 | 0.09 |

Fig. 7. Confusion matrix using (a) t-MLP and (b) CNN methods in the IEEE 30-bus system

### 3) Comparison with State-of-the-Art Methods

In this section, the proposed method is compared with some of the existing methods using the EMTP data set. In the first step, we take a look at the typical |WTC|s spectrum of the five events from each class. The fourth column in Table I shows the |WTC|s at bus 6 for different events. As these figures show, the peak values have almost the same magnitude during different events occurrence, except for the lightning, which is a more significant event. Moreover, calculating the energies for the |WTC|s using the method in [46] for classes 1 to 5 results in 1434, 1347, 1438, 36305, and 1404 V, respectively. The level of the calculated wavelet energies does not significantly change in different events (except for lightning). This indicates that distinguishing the events from each other based on their peaks or |WTC|s energies (spatial feature) does not lead to a desirable outcome. Furthermore, it can be seen from the |WTC|s that events 1, 2 and 3 (i.e., line energization, cap bank energization, and fault) last for almost 10 ms, and therefore, they both have nearly the same duration. This also shows that considering the duration of the occurrence of events (temporal feature) cannot be a good indicator for distinguishing the events from each other. Because of the similarities of the events, the traditional threshold-based methods (e.g., decision tree (DT)) have difficulties for an accurate identification of the cause of the underlying EMTE.

To further validate the effectiveness of the proposed method, similar to the RTDS case studies, we compare the overall accuracy with those obtained by energy-based methods, i.e., autoencoder, and SVM augmented with principal component analysis (PCA) [21], as well as t-MLP neural network.

To have a better insight into the classifiers' performance, four additional classification metrics other than accuracy (ACC) are calculated as well. These metrics are: precision (PRE), recall (REC), F1 score (F1), and false positive rate (FPR). These metrics are defined for a binary classification problem as follows:

$$PRE = TP/(TP + FP) \quad (4)$$
$$REC = TP/(TP + FN) \quad (5)$$
$$F_1 = 2 \times (PRE \times REC)/(PRE + REC) \quad (6)$$
$$FPR = FP/(FP + TN) \quad (7)$$

where TP is True Positive, which is the number of events that are correctly predicted to fall into the target class; FP is False Positive, which is the number of events that are incorrectly predicted to fall into the target class; FN is False Negative, which is the number of events that are incorrectly predicted to fall out of the target class; and TN is True Negative, which is the number of events that are correctly predicted to fall out of the target class.

In our multiclass problem, the metrics are still the same as the ones used in the binary classification. However, the metrics are calculated for each class by treating it as a binary classification problem after combining all non-target classes into the second class. Then, the binary metrics are averaged over all the classes to get either a macro average (treat each class equally) or micro average (weighted by class frequency) metric [36].

Table IV shows the results of four different methods in the IEEE 30-bus system. Based on the RTDS case studies, the input data is selected as 2D+W which outperforms the other methods. These results validate the superiority of the proposed CNN method that finds and uses the spatiotemporal correlation of the stream of voltage measurements from different locations. In contrast, the energy-based methods disregard the spatiotemporal information as the energy of the signals in several, or the entire sampling interval is calculated and utilized.

Fig. 7. (a) and (b) show the confusion matrix using the t-MLP and CNN methods for the IEEE 30-bus system where 80% of the data set is used for training and the remaining ones are used for evaluation. The training is performed for 4 epochs corresponding to 716 iterations. The confusion matrix shows the performance of the proposed cause identification method for distinguishing the correct events versus the misidentified ones. The rows shows the predicted class (Output Class) and the columns correspond to the actual class (Target Class). The diagonal cells correspond to events that are correctly predicted, and the off-diagonal cells correspond to events that are incorrectly predicted. Both the number of events and the percentage of the total number of events in each case are shown in each cell. The column on the far right of the confusion plot displays the precision and its error for each individual class. The row at the bottom of the plot displays recall and its error. Finally, the cell in the bottom right of the plot shows the overall accuracy and the overall error.

It can be seen that even though the t-MLP method is successful in distinguishing classes 1, 3, and 5, it completely fails to distinguish class 2 (cap energization) and class 4 (lightning) from other classes with 78.7% and 0% recall, respectively. However, our proposed method significantly outperforms the non-convolutional method with 100% and 96.7% recall in distinguishing class 2 and 4. Furthermore, it can be seen that the overall accuracy in the CNN outperforms the t-MLP with 99.7% against 92.4%.

## V. CONCLUSION

This paper presents a spatiotemporal unsupervised feature learning-based method for cause identification of electromagnetic transient events (EMTE) in the transmission system. The proposed method is validated to classify five EMTE as line energization, capacitor banks energization, faults, lightning, and high-impedance faults. As these events may not be easily classified by simply monitoring the relay voltage and current outputs or the peak values or duration of the time-frequency domain components of high-frequency voltage or current measurements, the proposed framework can be used in control rooms to increase the situational awareness. The proposed approach is based on the convolutional neural network (CNN) which is a combination of an unsupervised feature learning method and a supervised softmax classifier. The validation results show satisfactory performance of the proposed approach in comparison to the state-of-the-art events analysis and cause identification methods.